# Enhancement of the indistinguishability of single photon emitters coupled to photonic waveguides

**J. Guimbao[*], L.M. Weituschat, J.M. Llorens, P.A. Postigo**

*Instituto de Micro y Nanotecnología, IMN-CNM, CSIC (CEI UAM+CSIC) Isaac Newton, 8, E-28760, Tres Cantos, Madrid, Spain*
*[*]j.guimbao@csic.es*

**Abstract:** One of the main steps towards large-scale quantum photonics consists of the integration of single photon sources (SPS) with photonic integrated circuits (PICs). For that purpose, the PICs should offer an efficient light coupling and a high preservation of the indistinguishability of photons. Therefore, optimization of the indistinguishability through waveguide design is especially relevant. In this work we have developed an analytical model to calculate the coupling and the indistinguishability of an ideal point-source quantum emitter coupled to a photonic waveguide depending on source orientation and position. The model has been numerically evaluated through finite-difference time-domain (FDTD) simulations showing consistent results. The maximum coupling is achieved when the emitter is embedded in the center of the waveguide but somewhat surprisingly the maximum indistinguishability appears when the emitter is placed at the edge of the waveguide where the electric field is stronger due to the surface discontinuity.

## 1. Introduction

Indistinguishability of single photons generated by point defects is the central topic of quantum photonic integrated circuits for quantum information applications like quantum simulation [1], quantum teleportation [2] or quantum networks [3]. Indistinguishable photons are usually generated by parametric down-conversion [4] and lately from a single two-level quantum emitter in a solid-state environment [5]. Over the last years several on-chip integration of different SPS material systems have been demonstrated: III-V quantum dots [6], carbon nanotubes [7], NV [8] or SiV centers in diamond [9] and 2D layered materials [10]. For most of those solid-state quantum emitters the intrinsic indistinguishability at room temperature is almost zero because pure dephasing rates are orders of magnitude larger than the population decay rate [11]. Improvement of the indistinguishability can be achieved by low temperature operation and by reducing the radiative lifetime of the SPS using an optical cavity that takes advantage of the Purcell effect [12]. The balance between dephasing and population decay rates varies significantly depending on the material system. Whereas for specific single self-assembled GaAs quantum dots the emission at low temperature can be radiative lifetime limited [13], defects in 2D materials can exhibit several orders of magnitude of difference between radiative decay rate and pure dephasing rates [14]. Purcell enhancement using photonic resonators permits on-chip control of light−matter interaction to enhance collection efficiency and generation of indistinguishable photons [15] that can be used for on-chip processing of quantum information [16-18]. Therefore it is important to explore the coupling of SPS to PICs and its effect on the indistinguishability. In this work we use an analytical treatment of light radiation from a point source placed at an arbitrary location and with arbitrary orientation on a waveguide. For the waveguide material we have selected low-index of refraction materials commonly used in silicon photonics ($SiO_2$, $Si_3N_4$, Si) besides other high index materials like $WeS_2$ or $WO_3$ [19]. It is worth to note that other specially designed nanomaterials with ultra high refractive index can be designed [20].

We explore how the position of the source and its orientation affects the coupling to the waveguide modes and the indistinguishability of the photons. We also explore how the dimensions of the waveguide impact the indistinguishability. We perform FDTD simulations to validate the analytical model and to calculate the Purcell effect. The results show remarkable differences depending on the orientation of the SPS and provide maximum indistinguishability when the source is placed at the edge of the waveguide, in contrast to the maximum coupling efficiency position at the center of the waveguide. The indistinguishability is expressed in terms of the pure dephasing value of the emitter, so that the effects of the waveguide can be compared between strong and weak dissipative emitters. Depending on the waveguide geometry and the position of the source the indistinguishability can either increase or decrease, showing non-negligible enhancements for weak dissipative emitters placed at optimum positions.

Several works deal with the radiation of a point source embedded in bounded dielectric slabs and square waveguides through Green´s function methods [21-26]. Also, the problem of the unbounded dielectric slab is treated in [27] from a classical perspective and in [28] from a quantum perspective. However, in those cases the description of the source comes from the macroscopic expression of the dipole moment, without computing the Green´s Dyadic. The Green´s function of the unbounded 2D dielectric slab is covered in [29] and the same for the 3D cylindrical fiber in [30-32] through the development of a transform theory. As far as we know, the Green´s Dyadic of a 3D unbounded rectangular waveguide hasn´t been treated until this work. Here we develop a generalization of the transform theory from the 2D case [29] to obtain the solution of the 3D version of the problem for an unbounded rectangular waveguide. The obtention of the Green´s Dyadic allows us to directly connect the value of the indistinguishability with the geometrical parameters of the waveguide, which hasn´t been covered neither in all previously mentioned works.

## 2. Methods, results and discussion

*2.1 Indistinguishability for different SPS*

In an isolated two-level system the emission rate can be fully described by its population decay rate $\Gamma_0$. However, a solid-state quantum has an interaction with the mesoscopic environment. The two-level system is affected by random fluctuations of its energy that can be described by a stationary stochastic process characterized by a dephasing rate $\Gamma^*$ [33]. In this situation the indistinguishability (*I*) is reduced to [34]:

$$I = \frac{\Gamma_0}{\Gamma_0+\Gamma^*}, \tag{1}$$

In general, for any practical implementation in quantum information processing $I \geq 0.5$ [33]. The pure dephasing rates at room temperature of solid-state quantum emitters like color centers, quantum dots or organic molecules are about 3 to 6 orders of magnitude larger than their radiative decay rates [34]. Improvement of this efficiency can be achieved by working at cryogenic temperatures. For example, for excitons weakly confined in GaAs quantum dots the dot ground-state transition at low temperature is near radiative life-time limited [13] which would provide a balance of about $\Gamma^*/\Gamma_0 \approx 1$ and I ≈ 0.5. There are recent reports of even better performance with strain free GaAs/AlGaAs quantum dots without the need of Purcell enhancement [35]. For those highly efficient emitters the ratio $\Gamma^*/\Gamma_0 \to 0$ and the intrinsic indistinguishability tends to the unity. As an example of an intermediate situation, InAs quantum dots have decay and pure dephasing rates $\Gamma^*/\Gamma_0 = 2.6$ [36,37] and the indistinguishability is only *I*≈0.19. On the opposite side, strain-induced defects in 2D materials have typical radiative lifetimes in the order of nanoseconds with dephasing lifetimes in the order of picoseconds [14]. For those emitters the $\Gamma^*/\Gamma_0$ balance reduces to $10^{-3}$ with almost zero indistinguishability. However, recent works related to defects created in transition metal dichalcogenides (like $MoS_2$) by local helium ion irradiation [38] show radiative lifetimes <150 ps. Also, a lifetime <100 ps has been observed recently in regular strain induced defects in WSe$_2$ layers deposited on metallic surfaces [39, 40]. More examples of quantum emission demonstrations in 2D materials can be found in [41]. Therefore, emitters with a certain $\Gamma^*/\Gamma_0$ ratio may enhance significantly their indistinguishability when properly integrated inside photonic waveguides due to the change in their radiative decay rate. We will show that for certain geometries and emitter positions $I$ can be greatly reduced whereas optimal configurations can maintain or even enhance $I$ significantly, especially for emitters with a certain $\Gamma^*/\Gamma_0$ ratio.

*2.2 Analytic model for weak coupling*

We can assume that for a two-level emitter coupled to a waveguide the coupling ($g$) and the cavity decay rate ($\kappa$) are in the incoherent limit ($2g \ll \Gamma_0+\Gamma^*+\kappa$) and "bad cavity" regime ($\kappa \gg \Gamma_0+\Gamma^*$) [34]. In that limit the cavity can be adiabatically eliminated so the dynamics of the coupled system are described by an effective quantum emitter with decay rate ($\Gamma+R$) where $R$ is the is the population transfer between the emitter and the cavity [34]:

$$I = \frac{(\Gamma_0+R)}{(\Gamma_0+R)+\Gamma^*} \; ; \; R = \frac{4g^2}{\Gamma_0+\Gamma^*+\kappa} , \tag{2}$$

R is related to the Purcell enhancement ($P_f$) by $R = \Gamma_0 \cdot P_f$ [42]. Substituting in (2) we obtain:

$$I = \frac{(1+P_f)}{(1+P_f)+\frac{\Gamma^*}{\Gamma_0}}, \tag{3}$$

Here the Purcell enhancement is defined as $\Gamma/\Gamma_0$ where $\Gamma$ is the population decay rate in the inhomogeneous environment. This ratio is related to the power emitted by the source [43]:

$$\frac{\Gamma}{\Gamma_0} = \frac{P}{P_0}, \tag{4}$$

With $P$ and $P_0$ the power emitted in the inhomogeneous and homogeneous environment, respectively. The radiative decay rate enhancement can be obtained by FDTD simulations integrating the power emitted by the source inside the waveguide ($P$) and normalizing it respect to the power in a homogeneous surrounding ($P_0$). In order to extract the maximum amount of physical information from the interaction between the quantum emitter and the photonic waveguide, we develop an analytic model of the system. We use the relation between $\Gamma$ and the Green dyadic of the equation governing the interaction between the source and the waveguide. From (4) one can obtain the dependence of the decay rate with the imaginary part of the Green dyadic evaluated at the position of the source [43]:

$$\Gamma = \frac{4\omega^2}{\pi c^2 \hbar \varepsilon_0} [\mu \cdot Im\{\vec{G}(r_0,r_0)\} \cdot \mu], \tag{5}$$

Where $\omega$ is the frequency of emission of the source, $\varepsilon_0$ is the vacuum dielectric constant, c the speed of light in vacuum, $\hbar$ the reduced Planck constant, and $\mu$ the dipole moment of the source. Fig 1 shows a layout of a section of the waveguide used for our model. The waveguide (infinite in the *z*-axis) has a rectangular section filled with a linear homogeneous medium with refractive index $n_1$. The surrounding environment has a refractive index $n_2$=1.

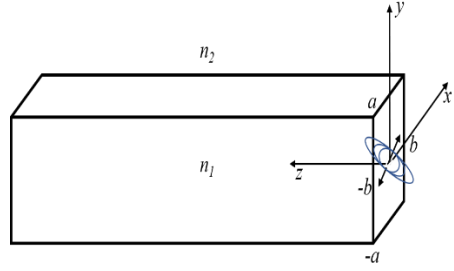

Fig. 1. Layout of the homogeneous infinite waveguide used for the analytical model

The calculation of the Green´s Dyadic is based on the development of a 3D transform theory applied to the unbounded Helmholtz equation. Details of the calculation and the explicit dependence with the waveguide width and the position/orientation of the source (for each contributing guided mode) can be found in the supplementary material. Using the Green dyadic we can obtain the Purcell enhancement as a function of the waveguide width for a point dipolar source that can be oriented parallel to the $x$-axis ($s$) or to the $y$-axis ($p$). The source is placed initially at the center of the waveguide cross-section ($x_0=0$, $y_0=0$). Initially, the waveguide thickness is arbitrarily fixed at $b=200$ nm and we will change the width ($a$) and the refractive index of the waveguide ($n_1$) using $n_1=1.44$, 2 and 3.4 corresponding to SiO2, SiN and Si respectively. This will provide some initial hints on how the system actually behaves.

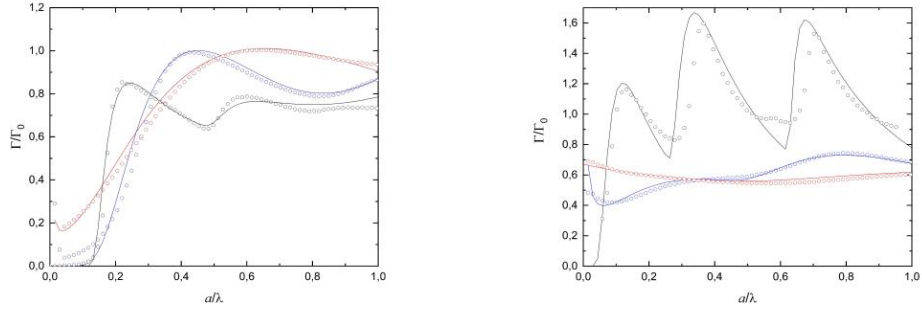

Fig. 2. Purcell enhancement of the radiative decay rate as a function of the wavelength-normalized waveguide width obtained from analytical calculations (lines) and FDTD simulations (open dots). $n_1=3.4$ (black), $n_1=2$ (blue), $n_1=1.44$ (red). (Left) Source orientation parallel to $x$-axis ($s$) (Right) Parallel to $y$-axis ($p$)

Fig. 2 shows the value of the Purcell enhancement, $\Gamma/\Gamma_0$, as a function of the normalized waveguide width, $a/\lambda$, for the before mentioned values of $n_1$ (1.44, 2, 3.4). Solid lines show the calculation of $\Gamma/\Gamma_0$ using (5) and open dots show the values obtained through FDTD simulations. Details of the FDTD simulations can be found in the supplementary material. Left panel in Fig. 2 shows the Purcell enhancement for the $s$-source. In general, it almost vanishes before the width reaches the cut-off of the $TE_{10}$ mode, which happens for $a/\lambda=0.13$, 0.1 and 0.05 for $n_1=3.4$, 2 and 1.44, respectively. Since the cut-off increases with $n_1$, the vanishing threshold also increases with $n_1$. After the cut-off, for increasing $a/\lambda$, The Purcell enhancement increases as the propagation constant decreases (with $1/a$) and the mode gets more confined. The maximum values for $\Gamma/\Gamma_0$ are 0.83, 1 and 1 when $a/\lambda=0.23$, 0.42 and 0.64 and the light confinement is maximum. If the waveguide becomes wider the modes spread out with lower intensity at the position of the source producing a decrease in $\Gamma/\Gamma_0$ that scales with $1/a$, until the cut-off with the second order mode is reached at $a/\lambda=0.43$, 0.8 and 1.2, respectively for the mentioned values of $n_1$. We note that there is no contribution from the lowest order $TE_{00}$ and

TM$_{00}$ modes because the components of the Green dyadic vanish at the position of the source for this orientation. This is expected since the *x*-components of the fundamental modes are antisymmetric respect to the source when it is placed at the center. For the *p*-source (Fig. 2, right panel) the situation is somewhat opposite and the components do not vanish at the position of the source for the lowest order TM$_{00}$. Since $b=200$ nm, in the case of $n_1=2$ and $n_1=1.44$ the cut-off condition is already reached at $a/\lambda =0$. For $n_1=3.4$ the cut-off is reached at $a/\lambda =0.05$. The Purcell enhancement follows a similar trend than for the *s*-source and shows maximum values of $\Gamma/\Gamma_0=1.2, 0.51$ and $0.6$ when $a/\lambda=0.13, 0.27$ and $1$ for the same values of $n_1$ than before. For both *s* and *p* orientations the Purcell enhancement exponentially decays with the width, although in a different trend due to the different $(m,n)$ values for each contributing mode. The maximum values for $\Gamma/\Gamma_0$ for the *s*-source are about 40% higher than for the *p*-source for $n_1=2$ and $n_1=1.44$. The reason is the value of the transverse electric field component of the TE$_{10}$, which is higher than the TM$_{00}$ at the position of the source ($x_0=0, y_0=0$) [44]. Nevertheless, for $n_1=3.4$ the maximum $\Gamma/\Gamma_0$ is about 40% higher for the *p*-source. This happens because for $n_1=2$ and $n_1=1.44$ the TE$_{10}$ mode is well confined for $b=200$ nm, but in the case of $n_1=3.14$ the TE$_{10}$ mode is not optimally confined, so the source has a better overlap with the TM$_{00}$. Therefore, high modal confinement and good spatial overlapping of the source with waveguide modes are key ingredients for Purcell enhancement, as one could intuitively expect. The indistinguishability should show its maximum value when the Purcell enhancement is maximum, according to (3). We note than a deviation in the optimal width of about 20 nm can decrease the Purcell enhancement, and therefore the indistinguishability, about 10%.

Since the position of the emitter is very relevant, we explore now its effect keeping fixed the waveguide widths in $a/\lambda=0.23, 0.42$ and $0.64$ (for the $n_1$ values as before) and for the *s*-source. We change the position of the source along the *x*-axis, from the center of the waveguide ($x_0/a=0$) to far way the edge ($x_0/a>\pm 1$).

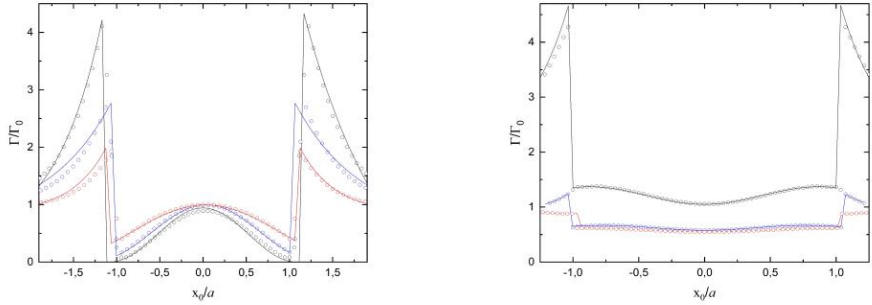

Fig. 3. Enhancement of the radiative decay rate as a function of the position of the point source (normalized respect to the waveguide width, *a*) obtained from the analytical model (lines) and from FDTD simulations (open dots). The origin in the *x*-axis corresponds to the center of the waveguide and the edges to $x_0/a=\pm1$. $n_1=3.4$ (black), $n_1=2$ (blue), $n_1=1.44$ (red). (Left) *s*-source (Right) *p*-source

Let´s focus first in the region inside the waveguide core ($x_0/a<\pm1$). Fig. 3 (left panel) shows the Purcell enhancement depending on the position of the *s*-source. As the *s*-source is separated from the center the overlap to symmetric modes decreases and the enhancement decreases as the source approaches to the edge of the waveguide. Therefore the maximum enhancement happens at the center of the waveguide. A deviation from that optimal position of about $x_0/a=\pm0.5$ leads to a decrease of the Purcell enhancement of about 20%. The opposite behavior is obtained for a *p*-source (Fig. 3 right). In this case the minimum overlapping is obtained at the center of the waveguide and the maximum enhancement is for about $x_0/a=\pm0.75$, where the overlap with the antisymmetric modes is maximum. FDTD simulations provide a maximum value for the enhancement of 1.42 matching the analytical calculations within an error of 0.2% for the Purcell enhancement and 0.3% for $x_0$.

The variation of the *coupling efficiency* with the position of the source inside the waveguide follows a similar trend than the Purcell enhancement. Details of coupling definition and its calculation can be found in the supplementary material. Fig. 4 shows the coupling efficiency depending on the position of the source for both *s* and *p* orientations. At the center of the waveguide, the *s*-source achieves a maximum coupling of $P_c/P_0 = 0.88$, 0.6 and 0.25 for $n_1=3.4$, 2 and 1.44 respectively, where $P_c$ is the emitted power coupled to guided modes. As expected, the coupling decreases with decreasing $n_1$. When the *s*-source is separated from the center, the coupling to symmetric modes decreases. Again, the opposite behavior is obtained for the *p*-source, which shows minimum coupling at the center of the waveguide.

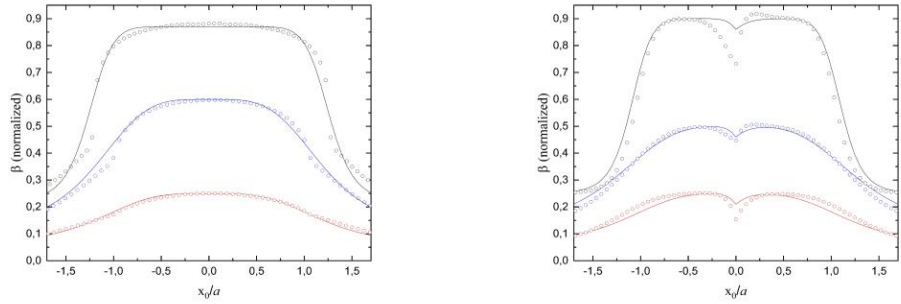

Fig. 4. Coupling efficiency versus normalized position of the source respect to waveguide width *a*. The origin in the *x*-axis corresponds to the center of the waveguide. The edges of the waveguide correspond to $x_0/a=\pm 1$. Figure shows results from analytical model (lines) and from FDTD simulations (open dots). $n_1=3.4$ (black), $n_1=2$ (blue), $n_1=1.44$ (red). (Left) *s*-source, (Right) *p*-source

Since there are recent experimental works that use heterogeneous integration of SPS and waveguides [45][46] it is worth to explore the dependence of the enhancement with the position of the source in the region outside the waveguide but close to its edge ($x_0/a>\pm 1$). Due to the index contrast between air and waveguide the electric field shows a strong discontinuity at the interface with an amount comparable to the square of the index ratio at the interface [47]. This effect can lead to a dramatic alteration of the mode profile in the vicinity of the edge that drastically increments the Purcell enhancement. When placed at the edge the enhancement is $\Gamma/\Gamma_0=4.2$, 2.6 and 1.9 for the *s*-source, and $\Gamma/\Gamma_0=4.6$, 1.2 and 0.87 for the *p*-source. The cost of this increase in the enhancement is a decrease in the coupling efficiency, which for the position at the edge is about $P_c/P_0 = 0.5$, 0.37 and 0.15 for the *s*-source and about $P_c/P_0 = 0.5$, 0.3 and 0.12 for the *p*-source. Details about this calculation can be seen in the supplementary material. At the points $x_0/a=\pm 1$ (i.e. the edges of the waveguide) the mode field shows its highest contrast according to $E_{clad} = (n_1/n_2)^2 E_{core}$ where $E_{core}$ is the field inside the waveguide and $E_{clad}$ is the field outside the waveguide. For that reason, the maximum value of the Purcell enhancement lies in the edges of the waveguide, especially for high $n_1$. Since the Purcell enhancement is strictly dependent on the field value at the position of the source, it´s maximum is achieved at the edge of the waveguide. On the other hand, the coupling is proportional to the guided-mode field value divided by the non-guided modes field value. Despite the guided-mode field value is maximum at the edge, the value of non-guided modes is also maximum at the edge. In consequence, the coupling at the edge is weaker than in the center of the core (where the coupling to non-guided modes is smaller).

Now that we have a better understanding of the physical meaning of the model we can explore simultaneously both degrees of freedom (i.e. $a$ and $b$) in order to find the optimal configurations in terms of the figures of merit. The source is placed initially at the center of the waveguide cross-section ($x_0=0$, $y_0=0$) in horizontal orientation (i.e. parallel to $x$-axis) but this time both $a$ and $b$ are varied from 0 to 0.7 $\lambda$. The results are obtained for five different values of the refractive index of the waveguide $n_1=1.44$, 2, 3.4, and 4.

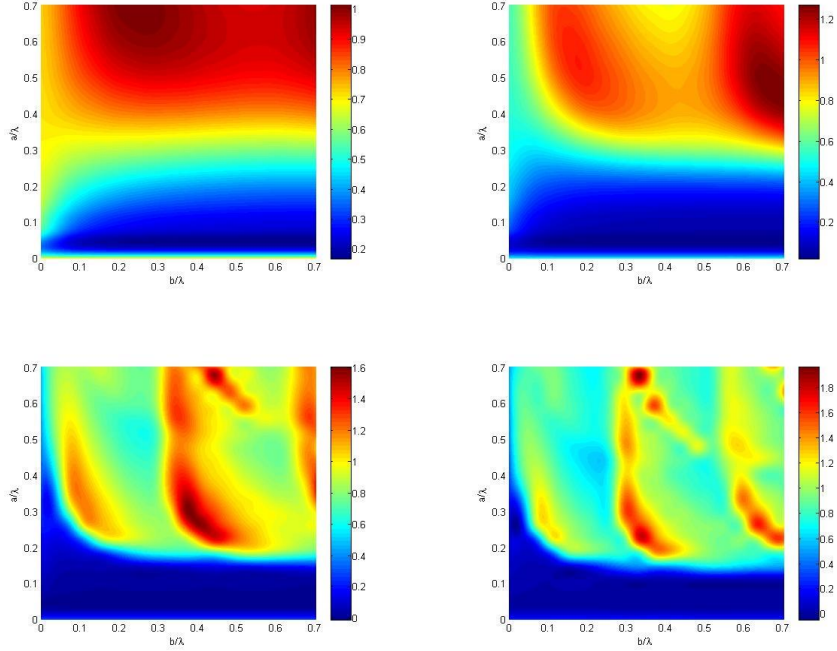

Fig. 5. Purcell enhancement as a function of the normalized width, $a/\lambda$, and thickness, $b/\lambda$, when the $p$-source is placed at the center of the waveguide. (Top-left) $n_1=1.44$ (Top-right) $n_1=2$ (Bottom-left) $n_1=3.4$ (Bottom-right) $n_1=4$.

Fig. 5 shows the value of $\Gamma/\Gamma_0$ as a function of the normalized waveguide width, $a/\lambda$, and normalized thickness, $b/\lambda$, calculated for the four different refractive indexes. The blue areas in the plots correspond to values of $a$ and $b$ below the first cut-off. The subsequent maxima and minima correspond to the activation of the TE$_{mn}$ and TM$_{mn}$ modes. For low refractive indexes (i.e. $n_1=1.44$, 2) the two first modes appear. As the refractive index increases the source starts to overlap effectively with the rest of higher order modes. The absolute maxima of $\Gamma/\Gamma_0$ increases with the refractive index, since the area of the spatial distribution of the modes decreases with $n_1$, so the field intensity gets higher at the position of the source. We obtain maximum values of $\Gamma/\Gamma_0=1$, 1.1, 1.6 and 1.9 for $n_1=1.44$, 2, 3.4 and 4, respectively. Due to the symmetry of the system the plots for the vertical source show the same rotated 90 degrees (see the supplementary material).

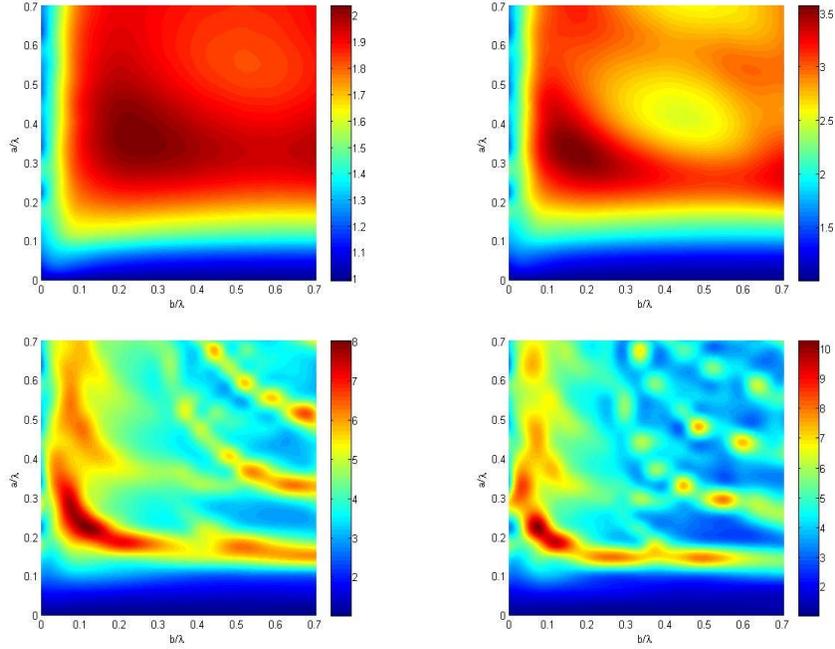

Fig. 6. Purcell enhancement as a function of $a/\lambda$ and $b/\lambda$ when the *p*-source is placed outside the core. (Top-left) $n_1$=1.44 (Top-right) $n_1$=2 (Bottom-left) $n_1$=3.4 (Bottom-right) $n_1$=4

Next, we place the source outside the waveguide, 10 nm away from the edge and oriented horizontally (i.e. parallel to *x*-axis). Fig. 6 shows the value of $\Gamma/\Gamma_0$ as a function of the normalized waveguide width, $a/\lambda$, and normalized thickness, $b/\lambda$, for different $n_1$. As we saw before, the field discontinuity generates a dramatic enhancement when the source is placed near the evanescent region of the mode. We observe that in all of the cases the maxima are located in the bottom left region, where both $a$ and $b$ have reached the cut-off but the first mode has not reached the maximum confinement. For that geometry, the mode is not optimally confined inside the core and the field gets accumulated at the edges of the waveguide so the overlap is more efficient. We obtain maximum values of $\Gamma/\Gamma_0$=2, 3.5, 8 and 10 for $n_1$=1.44, 2, 3.4 and 4 respectively. At this time the orientation of the source matters, since we can arrange two different configurations: (a) Parallel to the larger side of the waveguide (i.e. parallel to the *x*-axis if the source is placed on top of the core, or parallel to the *y*-axis if the source is placed on one side of the core); (b) Perpendicular to the larger side of the waveguide (i.e. parallel to the *y*-axis if the source is placed on top of the core, or parallel to the *x*-axis if the source is placed on one side of the core). The plots in Fig. 6 correspond to the second case. For the case when the source is oriented parallel to the surfaces we obtain lower maximum enhancements of $\Gamma/\Gamma_0$=0.9, 1.5, 6.8 and 7.1 for the different values of $n_1$.

The maximum enhancements obtained for the source at the edge can be used in (3) to obtain the maximum values for the indistinguishability. Fig. 7 shows *I* for an *s*-emitter placed at the edge of the waveguide versus the intrinsic emitter normalized dephasing ratio, $\Gamma^*/\Gamma_0$. Results for the *p*-source show an analogous behavior. From Fig. 7 we see that for *low* dissipative emitters with $\Gamma^*/\Gamma_0 \sim 1$ (like weakly confined GaAs dots of Ref. [13]) the expected indistinguishability can reach a value up to 0.8 when $n_1$=4, which makes an enhancement of *I* of about 30% in respect to the same dots without coupling to a waveguide. For InAs quantum dots with $\Gamma^*/\Gamma_0 = 2.6$ [36,37] we obtain $I \approx 0.6$, an enhancement of 40%. As the pure dephasing rate increases the indistinguishability decays asymptotically reaching 0.2 when $\Gamma^*/\Gamma_0 = 50$. Therefore, for strong dissipative systems with $\Gamma^*/\Gamma_0 > 50$ (like quantum emitters in 2D materials) the effect of the waveguide in the indistinguishability is very small. For emitters with lower dephasing ratio, $\Gamma^*/\Gamma_0 < 1$, and high intrinsic indistinguishability ($I >0.5$) the effect of the waveguide becomes again negligible since $I \rightarrow 1$ when $\Gamma^*/\Gamma_0 \rightarrow 0$. As $n_1$ increases the maximum $\Gamma^*/\Gamma_0$ for $I >0.5$ also increases reaching values up to 12 for $n_1=4$.

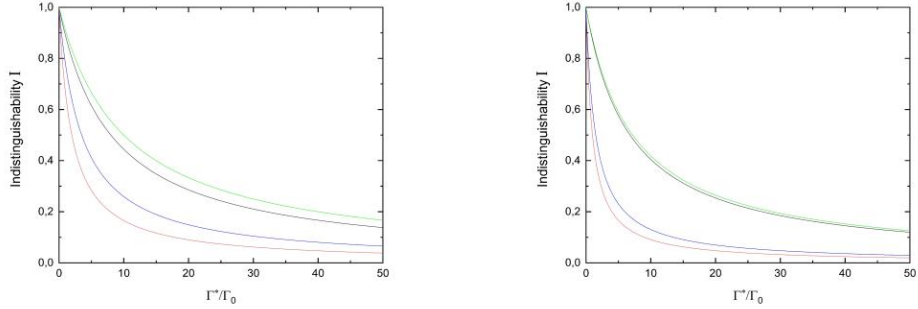

Fig. 7. Indistinguishability of an *s*-source quantum emitter placed at the edge of the waveguide versus its $\frac{\Gamma^*}{\Gamma_0}$ value. Green- $n_1$=4, Black– $n_1$=3.4, Blue- $n_1$=2, Red- $n_1$=1.44. (Left) *Perpendicular*-source (Right) *Parallel*-source

It is important to highlight that the positions for maximum indistinguishability differ from those of maximum coupling efficiency. Indistinguishability depends strongly on Purcell enhancement, which for the case of a waveguide achieves its maximum value at the edge, where the field is strongest. On the other side, the coupling efficiency on the edge is not as high as in the center, but still may have a value useful for some experiments or even some applications.

## 2.3 Analytic model for strong coupling

Finally, we explore the specific case of strong coupling in waveguides since it has been experimentally observed in many experiments [48-51] and also predicted to happen in 2D materials and waveguides [52]. For an emitter-waveguide in the strong-coupling regime ($2g \gg \Gamma_0 + \Gamma^* + \kappa$) the indistinguishability is [34]:

$$I = \frac{(\Gamma_0+\kappa)\left(\Gamma_0+\kappa+\frac{\Gamma^*}{2}\right)}{(\Gamma_0+\kappa+\Gamma^*)^2} ,  \qquad (6)$$

Since $\kappa = \omega/2Q$ and the quality factor is $Q = \omega/2(FHWM)$, where *FHWM* is the Full Width Half Maximum. We can compute $\kappa$ from the coupling of a source placed at the center of the waveguide to non-guided modes (i.e. free-radiation and evanescent modes) using the expansion of the Green dyadic in the continuous spectrum of solutions beyond the discrete subspace of guided modes [29] (see the supplementary material). We can compute the coupling to free-radiation modes as a function of the width using (7) and integrating the Pointing vector over a surface parallel to the XZ-plane at $z$=1 μm.

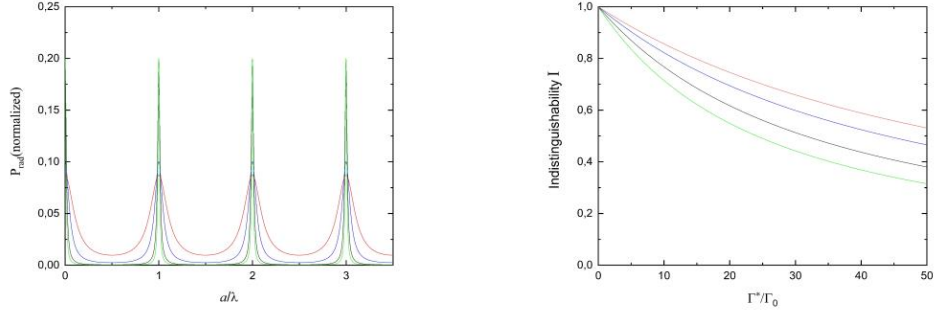

Fig. 8. (Left) Free radiation coupled to non-guided modes versus normalized waveguide width. The coupling is normalized with respect to power radiated by the source in a homogeneous environment. Green- $n_1$=4, Black– $n_1$=3.4, Blue- $n_1$=2, Red- $n_1$=1.44. (Right) Indistinguishability for a strongly coupled *s*-source quantum emitter placed at the center of the waveguide versus its $\Gamma^*/\Gamma_0$ value

Left panel in Fig. 8 shows the radiation coupled to non-guided modes versus the normalized width of the waveguide for $n_1$=4, 3.4, 2 and 1.44. As expected, the reflectivity of the effective mirrors of the resonator increases with $n_1$, increasing the Q-factor. We obtain a cavity decay rate of $\kappa = 504\ ps^{-1}, 557\ ps^{-1}, 730\ ps^{-1}$ and $1330\ ps^{-1}$ for $n_1$=4, 3.4, 2, 1.44 respectively. Using these values in (6) provides the indistinguishability. Right panel in Fig. 8 shows *I* for an *s*-emitter placed at the center of the waveguide for the previously obtained decay rates. For dissipative emitters with $\Gamma^*/\Gamma_0 \sim 1$ the expected indistinguishability can reach a value up to 0.81. As the dephasing ratio increases the indistinguishability decays smoothly. And as the index $n_1$ decreases the indistinguishability increases, because in the strong coupling regime the indistinguishability decreases with the effective time the photon is trapped in the cavity. Therefore, in the strong coupling regime, the increase in the waveguide Q-factor decreases the indistinguishability, which is in agreement with previous results [31].

## 3. Conclusions

We have calculated the indistinguishability of a point-source quantum emitter coupled to a waveguide because its technological implications in future quantum photonic integrated circuits. The emitter has arbitrary orientation and location respect to the waveguide. We have obtained the results for different index of refraction of the waveguide ($SiO_2$, $Si_3N_4$, Si, and other high index materials like $WeS_2$ or $WO_3$). The analytical model used permits a fast computing of the indistinguishability from a set of simple expressions derived from the same solution of the dyadic Helmholtz equation. The model has been numerically evaluated through 3D-FDTD simulations with excellent agreement. Maximum indistinguishability for an optimal waveguide width is found for a source placed outside the core, at the edge of the waveguide, in contrast to maximum coupling efficiency position at the center of the waveguide. For strong dissipative emitters with $\Gamma^*/\Gamma_0 > 50$ (like transition metal dichalcogenides) the effects of the waveguide in the indistinguishability are negligible but for low dissipative emitters with $\Gamma^*/\Gamma_0 \approx 1$ (like GaAs quantum dots) the indistinguishability can be enhanced up to a 30% and reach values around *I*≈0.8 when dots are coupled to a waveguide. We hope this work can help for an optimized design of PIC waveguides in quantum photonic circuits.

*Acknowledgments*

We gratefully acknowledge financial support from the European Union's Horizon 2020 research and innovation program under grant agreement No. 820423 (S2QUIP).

**References**

1. Aspuru-Guzik, A., & Walther, P. (2012). Photonic quantum simulators. Nature physics, 8(4), 285-291.
2. D. Fattal, E. Diamanti, K. Inoue, and Y. Yamamoto, Phys. Rev. Lett. 92, 037904 (2004).
3. H. Kimble, Nature (London) 453, 1023 (2008).
4. M. Eisaman, J. Fan, A. Migdall, and S. Polyakov, Rev. Sci. Instrum. 82, 071101 (2011).
5. Santori, C., Fattal, D., Vučković, J., Solomon, G. S., & Yamamoto, Y. (2002). Indistinguishable photons from a single-photon device. Nature, 419(6907), 594.
6. Chen, Y., Zhang, J., Zopf, M., Jung, K., Zhang, Y., Keil, R., ... & Schmidt, O. G. (2016). Wavelength-tunable entangled photons from silicon-integrated III–V quantum dots. Nature communications, 7(1), 1-7.
7. Khasminskaya, S. et al. Fully integrated quantum photonic circuit with an electrically driven light source. Nat. Photon.10, 727–733 (2016).
8. Mouradian, S. L., Schröder, T., Poitras, C. B., Li, L., Goldstein, J., Chen, E. H., ... & Lipson, M. (2015). Scalable integration of long-lived quantum memories into a photonic circuit. Physical Review X, 5(3), 031009.
9. Sipahigil, A., Evans, R. E., Sukachev, D. D., Burek, M. J., Borregaard, J., Bhaskar, M. K., ... & Camacho, R. M. (2016). An integrated diamond nanophotonics platform for quantum-optical networks. Science, 354(6314), 847-850.
10. Peyskens, F., Chakraborty, C., Muneeb, M., Van Thourhout, D., & Englund, D. (2019). Integration of single photon emitters in 2D layered materials with a silicon nitride photonic chip. Nature communications, 10(1), 1-7.
11. Kiraz, A., Ehrl, M., Hellerer, T., Müstecaplıoğlu, Ö. E., Bräuchle, C., & Zumbusch, A. (2005). Indistinguishable photons from a single molecule. Physical review letters, 94(22), 223602.
12. Gérard, J. M., Sermage, B., Gayral, B., Legrand, B., Costard, E., & Thierry-Mieg, V. (1998). Enhanced spontaneous emission by quantum boxes in a monolithic optical microcavity. Physical review letters, 81(5), 1110.
13. Gammon, D., Snow, E. S., Shanabrook, B. V., Katzer, D. S., & Park, D. (1996). Homogeneous linewidths in the optical spectrum of a single gallium arsenide quantum dot. Science, 273(5271), 87-90.
14. Aharonovich, I., Englund, D., & Toth, M. (2016). Solid-state single-photon emitters. Nature Photonics, 10(10), 631-641.
15. Dusanowski, Ł., Köck, D., Shin, E., Kwon, S. H., Schneider, C., & Höfling, S. (2020). Purcell enhanced and indistinguishable single-photon generation from quantum dots coupled to on-chip integrated ring resonators. Nano Letters.
16. Qiang, X., Zhou, X., Wang, J., Wilkes, C. M., Loke, T., O'Gara, S., ... & Wang, J. B. (2018). Large-scale silicon quantum photonics implementing arbitrary two-qubit processing. Nature photonics, 12(9), 534-539.
17. Wang, J., Sciarrino, F., Laing, A., & Thompson, M. G. (2019). Integrated photonic quantum technologies. Nature Photonics, 1-12.
18. Slussarenko, S., & Pryde, G. J. (2019). Photonic quantum information processing: A concise review. Applied Physics Reviews, 6(4), 041303.
19. Chen, C. T., Pedrini, J., Gaulding, E. A., Kastl, C., Calafiore, G., Dhuey, S., ... & Schwartzberg, A. M. (2019). Very high refractive index transition metal dichalcogenide photonic conformal coatings by conversion of ALD metal oxides. Scientific reports, 9(1), 1-9.
20. He, Y., He, S., Gao, J., & Yang, X. (2012). Nanoscale metamaterial optical waveguides with ultrahigh refractive indices. *JOSA B*, *29*(9), 2559-2566.
21. Huang, W., Yakovlev, A. B., Kishk, A. A., & Glisson, A. W. (2006, July). Dyadic Greens Function of the Hard Surface Rectangular Waveguide Verified Numerically by a Realistic Model. In *2006 IEEE Antennas and Propagation Society International Symposium* (pp. 2249-2252). IEEE.
22. Huang, W., Yakovlev, A. B., Kishk, A. A., Glisson, A. W., & Eshrah, I. A. (2005). Green's function analysis of an ideal hard surface rectangular waveguide. *Radio science*, *40*(05), 1-12.
23. Zhang, X., Xu, C., & Song, W. (2000, December). Calculating higher order mode characteristics of heteromorphic waveguide by operator theory. In *2000 Asia-Pacific Microwave Conference. Proceedings (Cat. No. 00TH8522)* (pp. 970-974). IEEE.
24. Słobodzian, P. M. (2002). On the dyadic Green's function in the source region embedded in waveguides or cavities filled with a stratified medium. *Microwave and Optical Technology Letters*, *35*(2), 93-97.
25. Klymko, V. A., Yakovlev, A. B., Eshrah, I. A., Kishk, A. A., & Glisson, A. W. (2005). Dyadic Green's function of an ideal hard surface circular waveguide with application to excitation and scattering problems. *Radio science*, *40*(3), 1-16.
26. Liu, S., Li, L. W., Leong, M. S., & Yeo, T. S. (2000). Rectangular conducting waveguide filled with uniaxial anisotropic media: A modal analysis and dyadic Green's function. *Progress In Electromagnetics Research*, *25*, 111-129.
27. Brueck, S. R. J. (2000). Radiation from a dipole embedded in a dielectric slab. IEEE Journal of Selected Topics in Quantum Electronics, 6(6), 899-910.
28. Creatore, C., & Andreani, L. C. (2008). Quantum theory of spontaneous emission in multilayer dielectric structures. Physical Review A, 78(6), 063825.
29. Santosa, F., & Magnanini, R. (2001). Wave propagation in a 2-D optical waveguide. SIAM Journal on Applied Mathematics, 61(4), 1237-1252.
30. Alexandrov, O., & Ciraolo, G. (2004). Wave propagation in a 3-D optical waveguide. *Mathematical Models and Methods in Applied Sciences*, *14*(06), 819-852.

31. Ciraolo, G., Gargano, F., & Sciacca, V. (2013). A computational method for the Helmholtz equation in unbounded domains based on the minimization of an integral functional. *Journal of Computational Physics*, *246*, 78-95.
32. Alexandrov, O. (2007). The far-field expansion of the Green's function in a 3-D optical waveguide. *Asymptotic Analysis*, *52*(1-2), 157-171.
33. Bylander, J., Robert-Philip, I., & Abram, I. (2003). Interference and correlation of two independent photons. The European Physical Journal D-Atomic, Molecular, Optical and Plasma Physics, 22(2), 295-301.
34. Grange, T., Hornecker, G., Hunger, D., Poizat, J. P., Gérard, J. M., Senellart, P., & Auffèves, A. (2015). Cavity-funneled generation of indistinguishable single photons from strongly dissipative quantum emitters. Physical review letters, 114(19), 193601.
35. Schöll, E., Hanschke, L., Schweickert, L., Zeuner, K. D., Reindl, M., Covre da Silva, S. F., ... & Rastelli, A. (2019). Resonance fluorescence of gaas quantum dots with near-unity photon indistinguishability. Nano letters, 19(4), 2404-2410.
36. J.M. Gérard, O. Cabrol, B. Sermage, Appl. Phys. Lett. 68, 3123 (1996).
37. P. Borri, W. Langbein, S. Schneider, U. Woggen, R.L. Sellin, D. Ouyang, D. Bimberg, Phys. Rev. Lett. 87, 157401-1 (2001).
38. Klein, J., Lorke, M., Florian, M., Sigger, F., Sigl, L., Rey, S., ... & Zimmermann, P. (2019). Site-selectively generated photon emitters in monolayer MoS 2 via local helium ion irradiation. Nature communications, 10(1), 1-8.
39. Chakraborty, C., Vamivakas, N., & Englund, D. (2019). Advances in quantum light emission from 2D materials. Nanophotonics, 8(11), 2017-2032.
40. Tripathi, L. N., Iff, O., Betzold, S., Dusanowski, Ł., Emmerling, M., Moon, K., ... & Schneider, C. (2018). Spontaneous emission enhancement in strain-induced WSe2 monolayer-based quantum light sources on metallic surfaces. ACS Photonics, 5(5), 1919-1926.
41. Chaudhary, R., Raghunathan, V., & Majumdar, K. (2020). Origin of selective enhancement of sharp defect emission lines in monolayer WSe2 on rough metal substrate. Journal of Applied Physics, 127(7), 073105.
42. Wein, S., Lauk, N., Ghobadi, R., & Simon, C. (2018). Feasibility of efficient room-temperature solid-state sources of indistinguishable single photons using ultrasmall mode volume cavities. Physical Review B, 97(20), 205418.
43. Novotny, L., & Hecht, B. (2012). Principles of nano-optics. Cambridge university press.
44. Chuang, S. L., & Chuang, S. L. (1995). Physics of optoelectronic devices.
45. Lombardi, P., Ovvyan, A. P., Pazzagli, S., Mazzamuto, G., Kewes, G., Neitzke, O., ... & Toninelli, C. (2017). Photostable molecules on chip: integrated sources of nonclassical light. ACS Photonics, 5(1), 126-132.
46. Errando-Herranz, C., Schöll, E., Laini, M., Gyger, S., Elshaari, A. W., Branny, A., ... & Bonato, C. (2020). On-chip single photon emission from a waveguide-coupled two-dimensional semiconductor. arXiv preprint arXiv:2002.07657.
47. Majumder, S., & Chakraborty, R. (2013). Semianalytical method to study silicon slot waveguides for optical sensing application. Optical Engineering, 52(10), 107102.
48. Brimont, C., Doyennette, L., Kreyder, G., Réveret, F., Disseix, P., Médard, F., ... & Alloing, B. (2020). Strong coupling of exciton-polaritons in a bulk GaN planar waveguide: quantifiying the Rabi splitting. arXiv preprint arXiv:2002.05066.
49. Oder, T. N., Lin, J. Y., & Jiang, H. X. (2001). Propagation properties of light in AlGaN/GaN quantum-well waveguides. Applied Physics Letters, 79(16), 2511-2513.
50. Solnyshkov, D. D., Terças, H., & Malpuech, G. (2014). Optical amplifier based on guided polaritons in GaN and ZnO. Applied Physics Letters, 105(23), 231102.
51. Ciers, J., Roch, J. G., Carlin, J. F., Jacopin, G., Butté, R., & Grandjean, N. (2017). Propagating polaritons in III-nitride slab waveguides. Physical Review Applied, 7(3), 034019.
52. Walker, P.M., Whittaker, C.E., Skryabin, D.V. et al. Spatiotemporal continuum generation in polariton waveguides. Light Sci Appl 8, 6 (2019).

**Supplementary material**

Our analytical model uses the explicit representation of the Green´s Dyadic of an arbitrary oriented 3D-point source in an unbounded infinite 3D-rectangular waveguide filled with a linear homogeneous medium of index $n_1$, and surrounded by a homogeneous medium of index $n_2$, (see Fig. 1 in the main text). We develop a partial eigenfunction expansion of the magnetic potential Green´s Dyadic involving the complete set of guided and non-guided eigenfunctions applying the transform theory from [1] to a 2D optical waveguide. Here we obtain a solution for the extended 3D problem of a rectangular waveguide by applying separation of variables on the transverse Laplacian operator. The expressions obtained allow the separation of the field energies into each of the guided modes and provides their dependence with the width of the waveguide and the position and orientation of the source. Since we explore the orientation of the dipole the effect of the waveguide thickness is an identical problem due to the symmetry of the system. Instead taking the dipole radiation analytical formula as the starting point for computing the Hertz-vector [2,10], we calculate the Green function because in this way we implement the effect of the near-field of the point-source [11]. The electromagnetic description of a quantum emitter (two level system) can be given by the dipole source approximation introduced in the Maxwell equations [11]. The approximation considers that the emission wavelength is several orders of magnitude larger than the size of the source. The lowest order of the quadrupole expansion for the distribution of the current has the following form [12]:

$$j(r) = -i\omega\mu\delta[r - r_0] \tag{S1}$$

In our model the domain under consideration is all $\mathbb{R}^2$ and each diagonal component of the magnetic potential vector $\overleftrightarrow{G}_A$ satisfies the scalar Helmholtz equation [13]:

$$\nabla^2 G_{A\upsilon\upsilon}(r, r') + n(x, y)k^2 G_{A\upsilon\upsilon}(r, r') = -\delta(r - r') \tag{S2}$$

Here $k$ is the time-frequency, $n(x, y)$ is he index of refraction, and $\upsilon = x, y, z$. The point-source is represented by the term $\delta(r - r')$. The core of the waveguide has an index $n_1$ with a width $a$ and a thickness $b$, and the cladding index is $n_2$. This is represented by:

$$n(x, y) = \begin{cases} n_1, & |x| < a \,;\, |y| < b \\ n_2, & |x| > a \,;\, |y| > b \end{cases} \tag{S3}$$

We recall the relevant results of [1], since we will follow a similar procedure the rest of the appendix. The associated eigenvalue problem is given by the homogeneous solutions of Equation (S2):

$$v_\upsilon(x, y, \beta)\exp(ik\beta z) \tag{S4}$$

Where each component $v_\upsilon(x, y, \beta)$ satisfies:

$$\nabla^2 v_\upsilon + [n(x, y)^2 - \beta^2]k^2 v_\upsilon = 0 \tag{S5}$$

Defining $q(x, y) = k^2 n(x, y)^2$ ; $\lambda = \beta^2 k^2$ and $d = k^2(n_1^2 - n_2^2)$ we get the eigenvalue equation:

$$\nabla^2 v_\upsilon + [\lambda - q]v_\upsilon = 0 \tag{S6}$$

With $Q = \sqrt{\lambda - d^2}$ there are two linearly independent solutions for each component: $v_s^\upsilon$ and $v_a^\upsilon$ given by:

$$v_j^\upsilon(x, y, \lambda) =$$

$$= \begin{cases} \phi_j^\upsilon(a, b, \lambda)\cos Q(x-a)\cos Q(y-b) + \dfrac{\phi_j^{\upsilon'}(a, b, \lambda)}{Q}\sin Q(x-a)\sin Q(y-b), & (x, y) > (a, b) \\ \phi_j^\upsilon(x, y, \lambda), & (|x|, |y|) < (a, b) \\ \phi_j^\upsilon(-a, -b, \lambda)\cos Q(x+a)\cos Q(y+b) + \dfrac{\phi_j^{\upsilon'}(a, b, \lambda)}{Q}\sin Q(x+a)\sin Q(y+b), & (x, y) < (a, b) \end{cases}$$

$$\tag{S7}$$

With $j \in \{s, a\}$. Each component $\upsilon$ from $\phi_j^\upsilon$ solutions satisfy different boundary conditions in the interval $(-a, a), (-b, b)$ [14]:

$$\frac{\partial \phi_s^x}{\partial x} = 0;\ \phi_s^y = 0;\ \phi_s^z = 0\ at\ x = -a, a$$

$$\phi_a^x = 0;\ \frac{\partial \phi_a^y}{\partial x} = 0;\ \frac{\partial \phi_a^z}{\partial x} = 0\ at\ x = -a, a$$

$$\phi_s^x = 0; \frac{\partial \phi_s^y}{\partial y} = 0; \phi_s^z = 0 \ at \ y = -b, b$$

$$\frac{\partial \phi_a^x}{\partial y} = 0; \phi_a^y = 0; \frac{\partial \phi_a^z}{\partial y} = 0 \ at \ y = -b, b$$

(S8)

For the bounded solutions with $0 < \lambda < d^2$ with corresponding eigenvalue $\lambda_{mn}^j$ we have [14]:

$$\phi_{smn}^x = \cos(m\pi x/a)\sin(n\pi y/b)$$

$$\phi_{smn}^y = \sin(m\pi x/a)\cos(n\pi y/b)$$

$$\phi_{smn}^z = \sin(m\pi x/a)\sin(n\pi y/b)$$

$$\phi_{amn}^x = \sin(m\pi x/a)\cos(n\pi y/b)$$

$$\phi_{amn}^y = \cos(m\pi x/a)\sin(n\pi y/b)$$

$$\phi_{amn}^z = \cos(m\pi x/a)\cos(n\pi y/b)$$

(S9)

The eigen-expansion of $G_{Avv}(r, r')$ can be separated into the guided part $G_{Avv}^g$ and free radiation part $G_{Avv}^{rad}$, which are given by:

$$G_{Avv}^{rad}(x, y, z, \xi, \zeta, \varsigma) = \sum_{j\epsilon\{s,a\}} \int_{d^2}^{\infty} \frac{e^{i|z-\varsigma|\sqrt{k^2 n_1^2 - \lambda}}}{2i\sqrt{k^2 n_1^2 - \lambda}} v_j^v(x, y, \lambda) v_j^v(\xi, \zeta, \lambda) \frac{\sigma_j^v(\lambda)}{\sqrt{\lambda - d^2}} d\lambda$$

(S10)

$$G_{Avv}^g(x, y, z, \xi, \zeta, \varsigma) = \sum_{j\epsilon\{s,a\}} \sum_{m=0} \sum_{n=0} \frac{e^{i|z-\varsigma|\sqrt{k^2 n_1^2 - \lambda}}}{2i\sqrt{k^2 n_1^2 - \lambda}} v_j^v(x, y, \lambda_{mn}^j) v_j^v(\xi, \zeta, \lambda_{mn}) r_{mn}^{jv}$$

(S11)

With $(\xi, \zeta, \varsigma)$ the position of the source, and $r_{mn}^{jv}$ and $\sigma_j^v(\lambda)$ defined as:

$$r_{mn}^{jv} = \left[\iint_{-\infty}^{+\infty} v_j^v(x, y, \lambda_{mn})^2 dx dy\right]^{-1} = \frac{\sqrt{d^2 - \lambda_{mn}^j}}{\sqrt{d^2 - \lambda_{mn}^j} \int_{-a}^{a} \int_{-b}^{b} \phi_{mn}^{jv} dx dy + \phi_{mn}^{jv}(a, b)^2}$$

(S12)

$$\sigma_j^v(\lambda) = \frac{\lambda - d^2}{(\lambda - d^2) \phi_j^v(a,b,\lambda)^2 + \phi_j^{v\prime}(a,b,\lambda)^2}$$

(S13)

The electric and magnetic Green dyadic ($\overleftrightarrow{G}_e$ and $\overleftrightarrow{G}_m$) are obtained from the potential magnetic vector $\overleftrightarrow{G}_A$ [51]:

$$\overleftrightarrow{G}_e(r,r') = \left(\overleftrightarrow{I} + \frac{1}{k^2}\nabla\nabla\right) \cdot \overleftrightarrow{G}_A(r,r') \tag{S14}$$

$$\overleftrightarrow{G}_m(r,r') = \nabla \times \overleftrightarrow{G}_A(r,r') \tag{S15}$$

Taking the $z$-component of the pointing vector, integrating over the cross section of the waveguide, and normalizing with respect to the emission of the source in the homogeneous case, we obtain a value for the coupling efficiency.

We have evaluated numerically the results of our analytical model through a series of FDTD simulations (Lumerical FDTD). The layout of the geometrical setup is shown in Fig. S1.

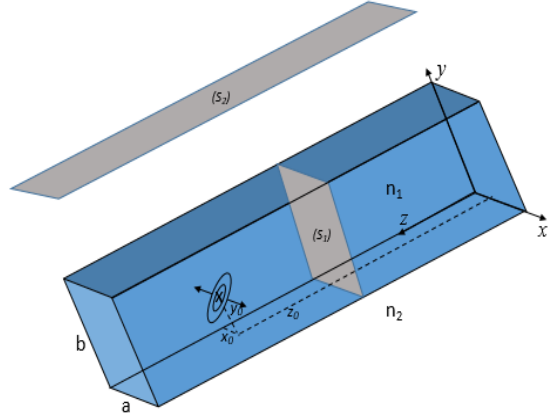

Fig. S1. Layout of the geometrical setup used for FDTD simulations. The dipole source is placed at ($x_0, y_0, z_0$) inside the core of a waveguide of length 3 $\mu m$, width $a$ and thickness $b$.

The waveguide is represented in the blue region with a thickness $b$ of 200 nm, a length of $3\mu m$ and index of refraction $n_1$=1.44, 2.0, 3.4, 4.0 and 4.5 at the wavelength of 750 nm. The numbers ($x_0, y_0, z_0$) represents the initial position of the source (an oscillating point charge) at the centre of the cross section and at a distance $z = 2.5\mu m$ from the origin. The source is oriented along the $x$-axis with the following emission parameters: pulse duration=100 ns, spectral width=10 MHz and central wavelength=750 nm. Planes PML boundary conditions are placed over a box of side 5 $\lambda$, and PML boundary conditions are set at the beginning and end of the waveguide. To obtain the coupling to guided modes we integrate the pointing vector Fourier transform over the surface $S_1$ with $xy$-dimensions equal to the waveguide thickness and width. We normalize with respect to the total emission in a homogeneous environment. The total integrated power of guided modes is multiplied by 2 to take in to account propagation in the two directions of the waveguide. In order to explore source positions outside of the waveguide, we choose PML boundary conditions over a square box of 3 $\mu m$ containing all the structure. For the coupling to non-guided modes we integrate over surface $S_2$ which is placed parallel to the $xz$-plane at a distance of 300 nm from the top of the core with a N.A.=0.55. The meshing in the region close to the source is set to $\lambda/100$ while for the rest of the structure is set to $\lambda/10$. Time resolution was 1 ns. For the computation of $\Gamma/\Gamma_0$ the Fourier transformed Pointing vector was integrated over the surfaces of a 10x10x10 nm squared box surrounding the source and then normalized with respect to the total emission in a homogeneous environment.

Due to the symmetry of the system, the results for $\Gamma/\Gamma_0$ varying both thickness and width the plots for the vertical source show the same as for the horizontal rotated 90 degrees:

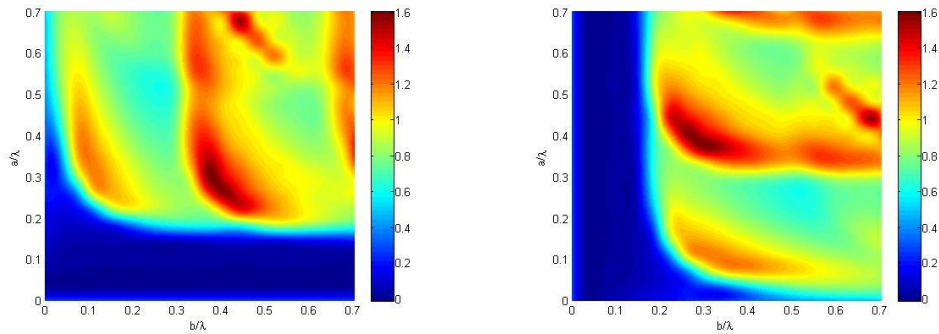

Fig. S2. Purcell enhancement of the radiative decay rate as a function of the wavelength-normalized waveguide width $a/\lambda$ and thickness $b/\lambda$, when the *p*-source is placed at the center of the waveguide for $n_1$=3.4. (Left) *s*-source. (Right) *p*-source.

**References**

1. Santosa, F., & Magnanini, R. (2001). Wave propagation in a 2-D optical waveguide. SIAM Journal on Applied Mathematics, 61(4), 1237-1252.
2. Brueck, S. R. J. (2000). Radiation from a dipole embedded in a dielectric slab. IEEE Journal of Selected Topics in Quantum Electronics, 6(6), 899-910.
3. Creatore, C., & Andreani, L. C. (2008). Quantum theory of spontaneous emission in multilayer dielectric structures. Physical Review A, 78(6), 063825.
4. Alexandrov, O., & Ciraolo, G. (2004). Wave propagation in a 3-D optical waveguide. Mathematical Models and Methods in Applied Sciences, 14(06), 819-852.
5. Bermel, P., Joannopoulos, J. D., Fink, Y., Lane, P. A., & Tapalian, C. (2004). Properties of radiating pointlike sources in cylindrical omnidirectionally reflecting waveguides. Physical Review B, 69(3), 035316.
6. Schneider, P. I., Srocka, N., Rodt, S., Zschiedrich, L., Reitzenstein, S., & Burger, S. (2018). Numerical optimization of the extraction efficiency of a quantum-dot based single-photon emitter into a single-mode fiber. Optics express, 26(7), 8479-8492.
7. Hoehne, T., Schnauber, P., Rodt, S., Reitzenstein, S., & Burger, S. (2019). Numerical Investigation of Light Emission from Quantum Dots Embedded into On-Chip, Low-Index-Contrast Optical Waveguides. Physica status solidI (b), 256(7), 1800437.
8. Verhart, N. R., Lepert, G., Billing, A. L., Hwang, J., & Hinds, E. A. (2014). Single dipole evanescently coupled to a multimode waveguide. Optics express, 22(16), 19633-19640.
9. Chen, Y., Nielsen, T. R., Gregersen, N., Lodahl, P., & Mørk, J. (2010). Finite-element modeling of spontaneous emission of a quantum emitter at nanoscale proximity to plasmonic waveguides. Physical Review B, 81(12), 125431.
10. Devaraj, V., Jang, Y., & Lee, D. (2016). Maximum photon extraction from a single quantum dot embedded in a metal/dielectric-cladded cylindrical structure. Journal of the Korean Physical Society, 68(8), 1014-1018.
11. Novotny, L., & Hecht, B. (2012). Principles of nano-optics. Cambridge university press.
12. De Wolf, D. A. (2001). Essentials of electromagnetics for engineering. Cambridge University Press.
13. Dyadic Green's Function: EECS 730 Winter 2009 c K. Sarabandi.
14. Hanson, G. W., & Yakovlev, A. B. (2013). Operator theory for electromagnetics: an introduction. Springer Science & Business Media.